\documentclass[aps,prl,twocolumn,showpacs,superscriptaddress]{revtex4-1}  

\pdfoutput=1
\usepackage{graphicx}
\usepackage{dcolumn}
\usepackage{bm}
\usepackage[T1]{fontenc}
\usepackage[latin9]{inputenc}
\usepackage{graphicx}
\usepackage{amsmath}
\usepackage{sidecap}
\usepackage{xfrac}

\renewcommand{\a}{\alpha}
\renewcommand{\b}{\beta}

\renewcommand{\d}{\delta}

\renewcommand{\r}{\mathbf{r}}

\newcommand{\hatt}{\mathbf{\hat{t}}}
\newcommand{\hatn}{\mathbf{\hat{n}}}
\newcommand{\hatb}{\mathbf{\hat{b}}}
\newcommand{\hatN}{\mathbf{\hat{N}}}
\newcommand{\hatT}{\mathbf{\hat{T}}}
\newcommand{\hatB}{\mathbf{\hat{B}}}
\newcommand{\Normal}{\mathbf{\hat{\mathcal{N}}}}
\newcommand{\pd}{\partial}

\newcommand{\gr}{\boldsymbol\gamma}

\newcommand{\bsubs}{\begin{subequations}}
\newcommand{\esubs}{\end{subequations}}

\newcommand{\be}{\begin{equation}}
\newcommand{\ee}{\end{equation}}

\newcommand{\bea}{\begin{eqnarray}}
\newcommand{\eea}{\end{eqnarray}}
\providecommand{\ba}{\begin{aligned}}
\providecommand{\ea}{\end{aligned}}

\newcommand{\mymat}[1]{\begin{pmatrix} #1 \end{pmatrix}}
\usepackage{amssymb}
\usepackage{multirow}

\newcommand{\R}{\mathbf{R}}

\begin{document}
\title{\textbf{Non-affine bending mode of thin cylindrical tubes}}
\date{\today}

\author{Efi Efrati}
\email{efi.efrati@weizmann.ac.il}
\affiliation{Department of Physics of Complex Systems, Weizmann Institute of Science, Rehovot 76100, Isreal}
\author{Yifan Wang}
\affiliation{Department of Physics, University of Chicago, 5720 S. Ellis Ave, Chicago, Illinois 60637, USA}
\affiliation{James Franck Institute, University of Chicago, 929 E. 57th St., Chicago, Illinois 60637, USA}
\author{Heinrich M. Jaeger}
\affiliation{Department of Physics, University of Chicago, 5720 S. Ellis Ave, Chicago, Illinois 60637, USA}
\affiliation{James Franck Institute, University of Chicago, 929 E. 57th St., Chicago, Illinois 60637, USA}
\author{Thomas A. Witten}
\affiliation{Department of Physics, University of Chicago, 5720 S. Ellis Ave, Chicago, Illinois 60637, USA}
\affiliation{James Franck Institute, University of Chicago, 929 E. 57th St., Chicago, Illinois 60637, USA}

\begin{abstract}
We discuss the response of thin cylindrical tubes to small indentations. The stretching-free embedding of these surfaces is completely determined by the embedding of one edge curve, thus reducing their bending response to an effectively one dimensional problem. We obtain that in general thin cylindrical tubes will bend non-uniformly in response to an indentation normal to their surface. As a consequence, their local bending stiffness, as measured by indenting at various locations along their axis, is predicted to peak at the center and decrease by a universal factor of 4 at either end of the tube. This characteristic spatial profile of the bending stiffness variation allows a direct determination  of the bending modulus of very thin tubes that are shorter than their pinch persistence length, $l_{p}\propto R^{3/2}/t^{1/2}$, where $R$ is the tube diameter and $t$ its wall thickness. We compare our results with experiments performed on rolled-up stainless steel sheets.

\end{abstract}

\maketitle

\begin{figure}[t]
\begin{center}
\includegraphics[width=7.5cm]{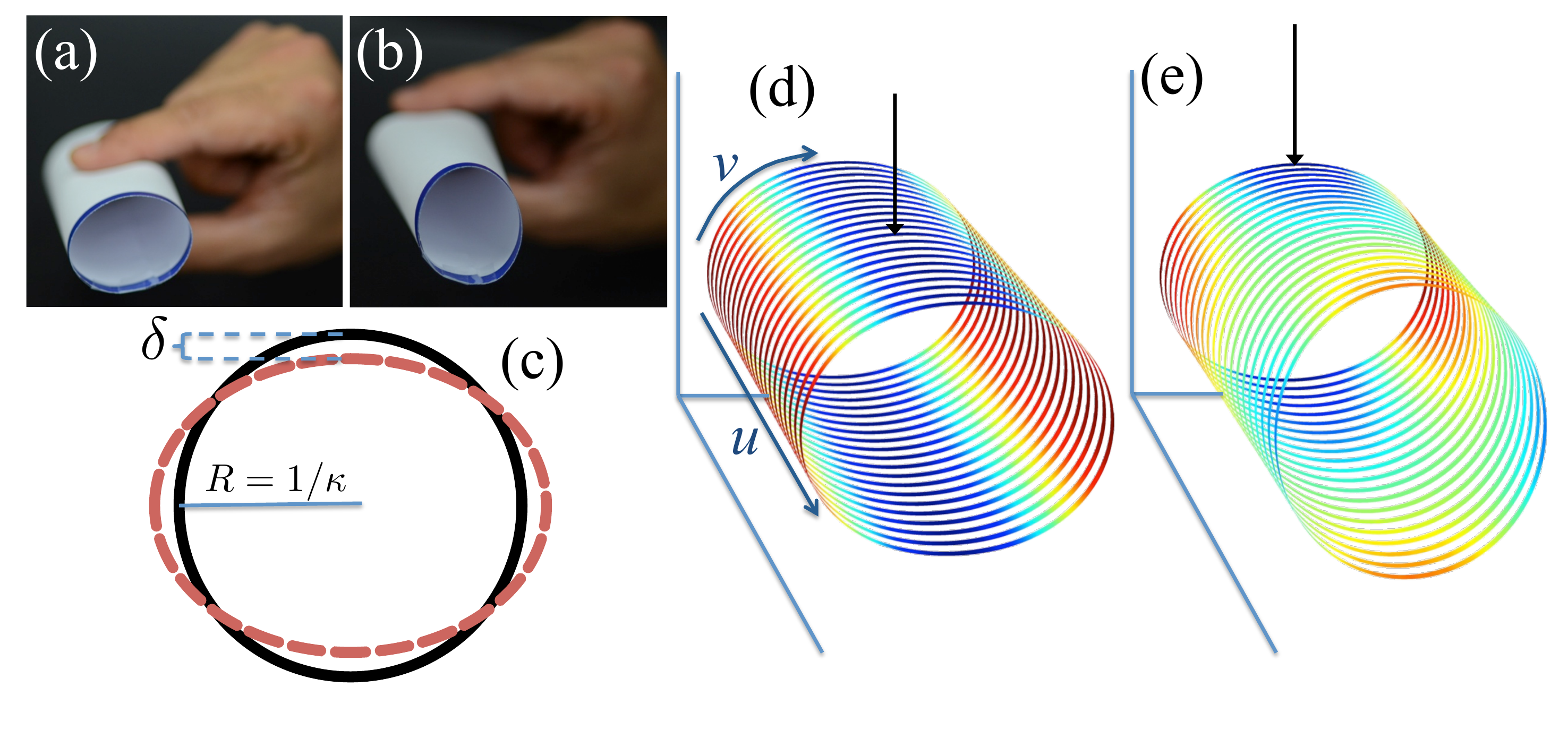}
\end{center}
\caption{(a) A thin cylindrical tube pinched at its center shows an affine bending response.(b) The same cylindrical tube pinched at the far end shows a non-affine response with the near end narrowing horizontally. (c) The geometry of deformation of a circle. (d) and (e) Tubes considered as a collection of rings with linearly varying perturbation amplitude as discussed in the text. Colors correspond to curvature increasing from blue to red, with green corresponding to the unperturbed value. Pushing the cylindrical tube at its center results in an affine response that contains significant areas of increased curvature, (d).  Pushing the cylindrical tube at its end results in a non-affine response that contains much less areas of increased curvature, (e). }
\label{fig:rings}
\end{figure}

If one rolls a sheet of paper it into a cylindrical tube of circular cross section and then pinches it slightly between two fingers at its center, the cylindrical tube will bend uniformly, keeping all cross-sections identical (in an affine fashion). If, however, the pinch is applied near one of the ends of the cylindrical tube, say oriented along the vertical direction, the cross-section on the far end of the tube will extend in the vertical direction and narrow along the horizontal direction (thus deforming the tube in a non-affine fashion), as depicted in Figure \ref{fig:rings}. We show that this mode of deformation can be carried out isometrically and only necessitates bending deformations. It is thus expected to be the dominant response to normal indentation in very thin cylinders. 

The elastic energy of thin elastic sheets is composed of two contributions associated with distinct modes of deformation. The in-plane stretching contribution to the energy, which scales linearly with the thickness of the sheet, penalizes lateral stretching and compression of the thin sheet. The out-of-plane bending contribution, which is cubic in the thickness of the sheet, penalizes the curving of the surface. A general deformation of a thin sheet will contain both types of deformations. However, for thin enough sheets bending deformations become energetically favorable compared with in plane stretching deformations. Thus, whenever a sufficiently thin sheet can accommodate an external constraint via a finite energy bending response with no stretching, it will do so. The externally induced deformation in this case could be accommodated by a pure bending mode and thus the resistance to the deformation will scale with the thickness cubed. However, in some cases no such configurations with finite bending and no stretching exist; the elastic energy associated with a pinch in a long pipe scales as thickness to the power of $5/2$ \cite{MVD07}, and the elastic energy of the ridge between two sharp conical defects scales as thickness to the power of $8/3$ \cite{Lob96}. In both cases the sub-cubic scaling implies 
that significant stretching and bending deformations contribute to the energy. 

Recent experiments on thin elastic sheets studying atomically thin graphene monolayers and nanoparticle membranes challenge the applicability of the physical thickness in determining the mechanical properties of the material \cite{WLMELJ15, BBR+15}. In these cases the bending stiffness and stretching stiffness must be measured independently in order to assess if the sheet is to be considered ``thin''. Moreover, even when simple elasticity theory holds and the scaling of stretching and bending stiffness with respect to the physical thickness is obeyed, it is often not possible nor practical to vary the thickness of a thin sheet to classify the type of deformation observed. 

In this work we show that the deformation described in Figure \ref{fig:rings} can be achieved without introducing any in-plane stretching to the system, and show how the minimization of bending energy favors a non-affine bending response, which in turn diminishes the bending energy by up to a factor of four. The non-affine response endows the cylindrical tube with variable resistance to normal loads depending on the location of load application along the tube. The characteristic shape of this resistance curve serves to verify that the cylindrical tube undergoes a pure bending deformation, and  allows the extraction of a bending stiffness without varying the effective thickness of the structure to verify the appropriate scaling for the bending rigidity.

These results can be understood by considering the Wunderlich function giving the bending energy of wide developable ribbons \cite{Wun62}, and a recent generalization of it \cite{DA14}. These works rely on the ability to completely determine the embedding of the developable surface by knowing of the embedding of its edge curve. It is important to note that if the deformed cylindrical tube is not thin enough (or alternatively very long), a localized deformation persisting only along a length $l_p$ near the loading location will be favorable. This type of deformation includes in-plane stretching and was discussed in \cite{MVD07,San13}.

While the fully non-linear equations of the exact isometry are solvable, we first present a linearized ansatz to provide intuition for the exact solution. We consider a simple deformation of a circle as depicted in Figure \ref{fig:rings}(c).
The amplitude of deformation is $\d$. As the perturbed circular configuration is of minimal bending energy, the leading term contributing to the bending energy is quadratic in $\d$, i.e.
\[
\int \kappa^2 ds= 2\pi\kappa_0+C \d^2+\mathcal{O}(\d^3).
\]
Now assume that a cylindrical tube is bent such that its unperturbed circular cross sections remain almost planar and the principal direction of curvature remains roughly along the circumference. In such a case we may consider the deformed cylindrical tube simply as a collection of rings, each associated with the energy above.
The non-trivial geometric constraint, that the surface remain developable, is enforced by letting $\d$ vary only linearly with the parameter $u$ along the axial direction:
\begin{equation}
E\approx const+  C \mathop{\int}_0^L(\d_0+ \b u)^2 du.
\label{eq:approxE}
\end{equation}
The above form sets the displacement at $u=0$ to the value $\d_0$. A uniform bending is associated with the energy cost of $C\d_0^2 L$. If, however, non-affine bending is allowed we obtain a minimum for the bending energy at $\b=-3 \d_0/(2L)$ at one quarter of the  energy cost of the affine bending, $C\d_0^2 L/4$.
We can change the integration limits to read $-\a L$ to $(1-\a)L$, with  $0\le\a\le1$. This corresponds to pushing the cylindrical tube down a distance $\a L$ from its boundary.
The minimal energy in this case reads
\begin{equation}
E\approx const+  \frac{C \d_0^2 L}{4} \frac{1}{1-3\a+3\a^2}.
\label{eq:Erings}
\end{equation}

When bending in an affine fashion the deviation from the minimal bending configuration is distributed uniformly. However, the non-affine bending mode allows this deviation from the minimal bending configuration to diminish away from the point at which the deformation is prescribed. It is therefore natural to expect that this type of deformation, whenever available, will be favorable energetically. We next prove that such a non-affine deformation can be performed without any in-plane stretching, and carry out the above calculation in a geometrically exact setting.

%
%
%
%

Let $\R(u,v)$ be the surface obtained by some isometric deformation of a cylindrical surface. The surface is developable and can therefore be given by
\begin{equation}
\R(u,v)=\gr(v)+u\hatT(v),
\label{eq:surf}
\end{equation}
where u is the axial co-ordinate and v is the azimuthal co-ordinate, and we have the freedom to set $ |\gr'|=1$ (henceforth prime denotes differentiation with respect to $v$).  For the case of a cylindrical tube at rest $\gr$ forms a circle in the $xy$ plane parameterized by arc-length parameter,$v$, and we have $\hatT=\hat{\mathbf{z}}$.

For the base curve we have the Serret-Frenet frame with $\gr'=\hatt$, $\gr''=\hatt'=\kappa \hatn$, and $\hatt\times\hatn=\hatb$, which satisfy
\[
\pd_v\mymat{\hatt\\ \hatn \\ \hatb}=\mymat{0&\kappa (v)&0\\-\kappa (v)&0&\tau (v)\\0&-\tau (v)&0}\mymat{\hatt\\ \hatn \\ \hatb},
\]
where $\kappa (v)$ and $\tau (v)$ are the standard curvature and torsion of the base curve. We also set
$\hatT'=K(v) \hatN$, and $\hatN'=-K(v)\hatT+\Omega (v) \hatB$, where $\hatB=\hatT\times\hatN$. Note that $K(v)$ ($\Omega (v)$) is NOT the curvature (torsion) of the axial $u$-parameter curves. These curves are straight and extend along the $u$ coordinate whereas prime here denotes differentiation along the $v$-coordinate.

With the aid of the above simplifications we may write the metric of the surface:
\begin{equation}
ds^{2}=du^{2}+2\cos(\theta)du\,dv + (1+u^2K^2+2 u K \sin(\theta))dv^{2},
\label{eq:metric}
\end{equation}
where $\cos(\theta)=\hatt\cdot\hatT=\gr'\cdot\hatT$. 
We can express the tangent to the base curve, $\hatt$, using the orthonormal triplet $\hatT$, $\hatN$ and $\hatB$. To do so we consider the second fundamental form $b_{ij}=\partial_{i}\partial_{j}\r\cdot\Normal$, with $\Normal$ the normal to the surface. Eq. \eqref{eq:surf} implies $b_{11}=0$. Vanishing of the Gaussian curvature ($|b|=0$) yields $b_{12}=0$, which in turn leads to $\hatt\cdot\hatB=0$. We may thus write
\begin{equation}
\gr'=\hatt=\cos(\theta) \hatT+\sin(\theta) \hatN.
\label{eq:hatt}
\end{equation}
Note that in particular the above expression implies that the surface's normal is given by $\Normal=\hatT\times\hatN=\hatB$, independently of $u$.
The only non vanishing component of the second fundamental form reads
\begin{equation}
b_{22}=\sin(\theta)\Omega+uK\Omega,
\label{eq:b}
\end{equation}
where above $\Omega=\sqrt{\kappa^{2}-k_g^{2}}/ \sin(\theta)$, as can be shown by differentiating equation \eqref{eq:hatt}. Further differentiation of the equations relating $\hatt,\hatn,\hatb$ and $\hatN,\hatT,\hatB$, and the vanishing of the geodesic curvature of the $u=0$ curve provide the closure relation for $K$ and $\theta$, expressing them in terms of base curve properties alone
\begin{equation}
0=-K-\theta',\qquad
\cot(\theta)=\frac{\tau}{\kappa}\equiv \eta.
\label{eq:k}
\end{equation}

Expressing all surface properties in terms of the base curve's curvature and torsion yields the well known Wunderlich functional \cite{Wun62,Tod14,DA14}
\begin{equation}
\begin{aligned}
E_B&=Y_B \int\!\!\int 4H^2 dA=Y_B \int\!\!\int \left(\frac{a_{11}b_{22}}{|a|}\right)^{2}\sqrt(|a|)du\,dv\\
&=
Y_B\int\frac{\kappa^2(1+\eta^2)^2}{\eta'} \log \left( \frac{1+(1-\a)L\eta'}{1-\a L\eta'}\right)ds,
\label{eq:Wun}
\end{aligned}
\end{equation}
A standard limit for the Wunderlich energy \eqref{eq:Wun} is that of a narrow ribbon, expanding in orders of $L$ as in Refs.\cite{Sad30,MK93,Efr15}. The first non-vanishing term, which gives rise to the Sadowsky functional  \cite{Sad30}, is linear in $L$ and does not depend on $\eta'$. In the present context we are interested in a different case, that of small perturbations of a long cylindrical tube, which in turn leads to the limit of a small perturbations of a circular curve. Based on the result obtained through the hand waving arguments above we assume that $\delta \kappa\propto \delta\eta$, and expand the energy to second order in the perturbation fields:
\[
\begin{aligned}
E_B\approx&Y_B\int \bigl( L (\kappa_0+\d \kappa)^2+L^2\kappa_0(2\a-1)\d\eta\, \d\kappa' \\
& +2 L\kappa_0^2 \d \eta^2+L^3\kappa_0^2(\a^2-\a+\tfrac{1}{3})\eta'^2
\bigr)ds.
\end{aligned}
\]
\begin{figure}[t]
\begin{center}
\includegraphics[width=7.5cm]{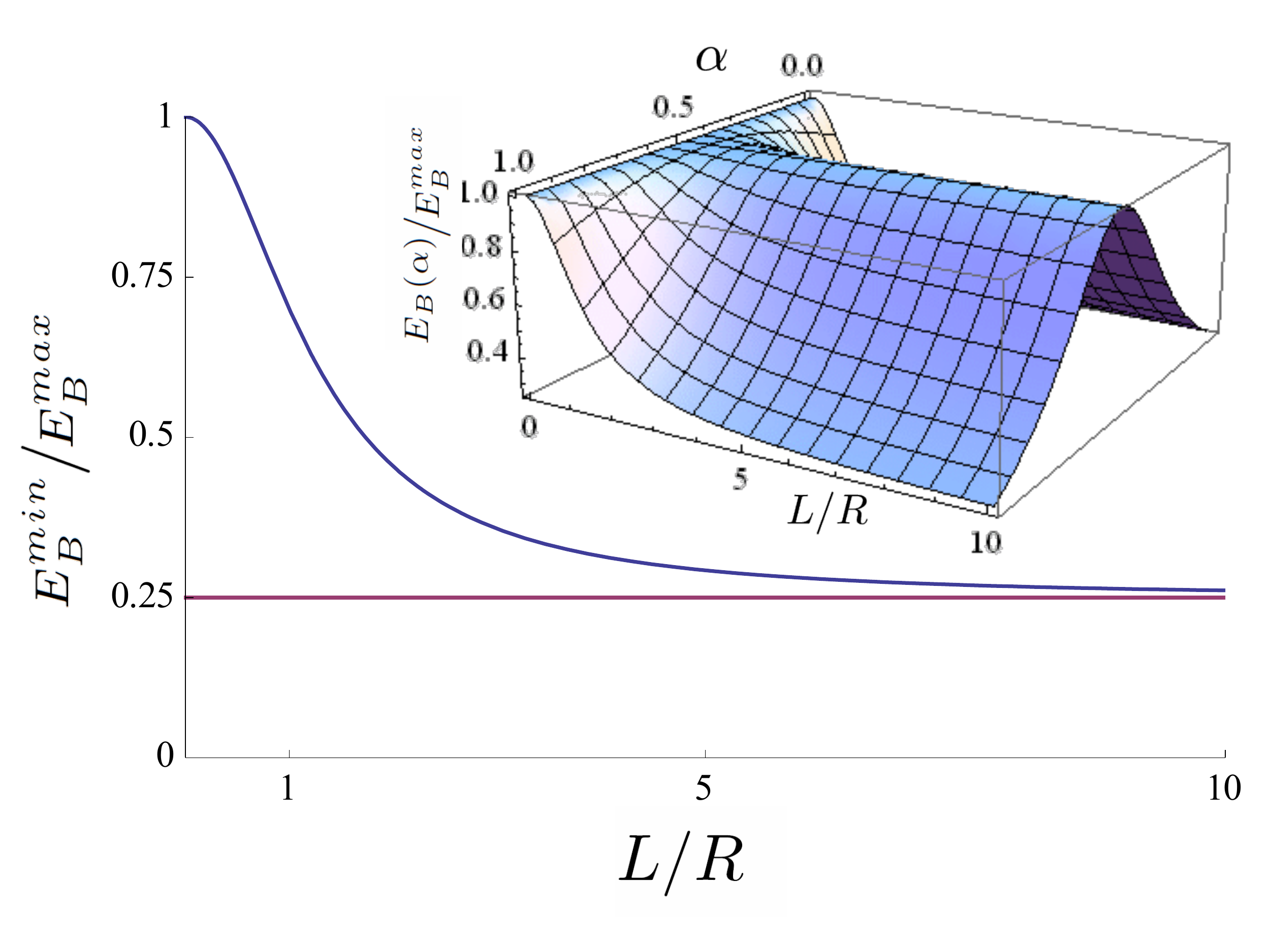}
\end{center}
\caption{Normalized stiffness. The inset shows the normalized stiffness as a function of cylindrical tube length $L$ and load application position. Note that the bell shaped stiffness profile flattens completely for short cylindrical tubes. The main figure shows the ratio of minimal stiffness to maximal stiffness, which asymptotes to the value $1/4$. }
\label{fig:2D}
\end{figure}
For a constant curvature (vanishing $\d\kappa$) the affine bending mode where $\d\eta=0$ is favorable. However, if the curvature is not constant the second term allows lowering the bending energy below that of the affine deformation by introducing torsion to the base-curve. Assuming a single mode response $\d\kappa= \epsilon \cos(2 s)$, with $\epsilon=-3\kappa_0^2 \d z$ where $\d z$ is the maximal normal displacement of the cylindrical tube along the radius,
 we obtain a perturbed elastic energy $E_B$ which we expand to second order in $\d z$. The stiffness of the cylindrical tube to normal perturbation is then given by
\begin{equation}
 \frac{d E_B}{d \d z}\equiv F_z = Y_B \frac{3 \pi L\kappa_0^3(6+L^2\kappa_0^2)}{1+2L^2\kappa_0^2(\tfrac{1}{3}-\a+\a^2)} \d z+\mathcal{O}(\d z)^2
 \label{eq:stiffness}
 \end{equation}
 For short cylindrical tubes the stiffness is expected to be independent of $\a$, whereas for long cylinders we recover the result of Equation \eqref{eq:Erings} where the ratio of maximal to minimal stiffness yielded $1/4$, as depicted in Figure \ref{fig:2D}. Recalling that for the present analysis to be valid the length of the cylinder must be smaller than the pinch persistence length \cite{MVD07}, the long cylinder limit should be employed cautiously.

\begin{figure}[t]
\begin{center}
\includegraphics[width=7.5cm]{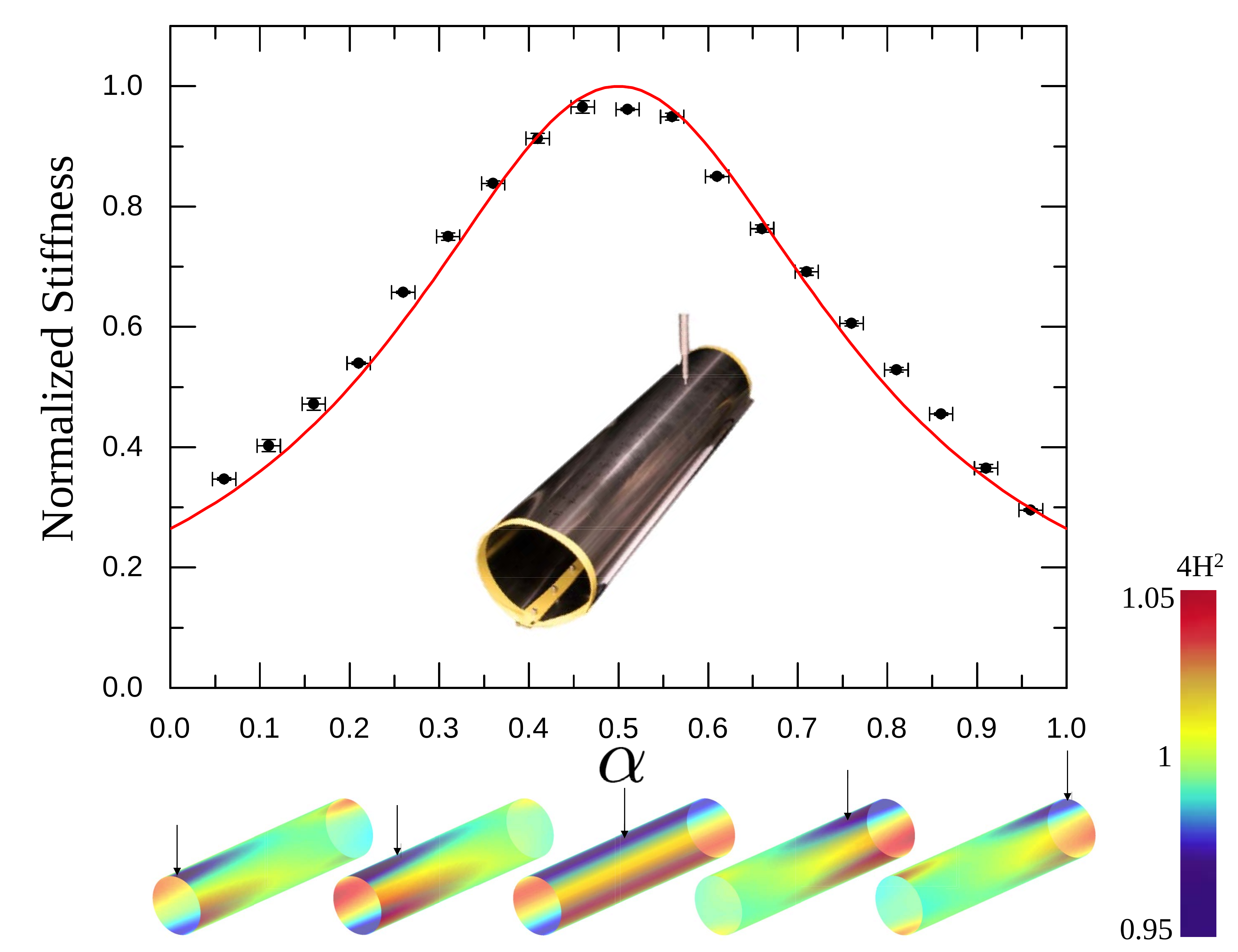}
\end{center}
\caption{Stiffness as a function of normalized load application position , $\alpha = x/L$, where $x$ is the lateral distance from the edge of the tube to the load application point and $L$ is the total length of the tube. Here the stiffness has been normalized by its peak value at $\alpha = 1/2$.
The measured stainless steel cylinder has a radius of $46$mm, a length $L=400$mm, and measured wall thickness of $0.10$mm. The red line represents the prediction from Eq. \eqref{eq:stiffness}. Colored graphs are Matlab calculation results of the cylinder deformation under indentation at different positions as shown by the arrows.}
\label{fig:metal}
\end{figure}

In order to compare this non-affine bending model with experimental results, a thin stainless steel cylinder was indented using an Instron 5869 materials tester, as shown in Figure \ref{fig:metal}. The characteristic bell shape of the stiffness profile conforms very well to the prediction from Equation \eqref{eq:stiffness} (red line), which verifies that indeed we observe the energy associated with the non-affine bending mode. In addition, from the absolute value of the stiffness we may extract the bending stiffness, and calculate the effective thickness \cite{Efr15},
\begin{equation}
 Y_B = \frac{Yt^3}{24(1-\nu^2)}.
\end{equation}
Here $Y=94GPa$ is the metal sheet's measured Young's Modulus and $\nu=0.33$ is the Poisson's ratio. From the measures value of the bending stiffness $Y_B=4.2 \times 10^{-3}J$ we may deduce the effective thickness to be $t=0.099$mm in good agreement with the measured thickness of $0.010$mm. For the metal cylinder depicted in Figure \ref{fig:metal} the persistence length measures over $90cm$, more than twice the length of the cylinder.

Our calculations show that the signature of the non-affine bending response is a bell-shaped response curve describing the stiffness (resistance to normal force application) as a function of location along the cylinder, which is four times as stiff at its center than near one of its ends. We demonstrate that whenever such a response curve is obtained, a bending stiffness can be directly extracted from it. To assure self-consistence one needs to examine the pinch persistence length $l_{p}\approx D^{3/2}\left(\frac{Yt}{Y_B}\right)$, where $Y_B$ is the obtained bending rigidity, the product $Yt$ represents the 2D stretching modulus and $D$ is the tube diameter \cite{MVD07}. The nearly stretching-free scenario discussed here is expected to hold provided that the length of the tube is much smaller than the persistence length
\footnote{Note that small bending stiffness (corresponding to small thickness) leads to large pinch persistence length and thus to a wider regime of applicability of the theory here. }.
If, however, the cylinder becomes longer than the persistence length of a pinch \cite{WLMELJ15}, then the minimal energy solution will include a non-negligible amount of stretching energy. In such a case, the present analysis based on isometric deformations would not hold.

Identifying the pure bending response described above is expected to be exceptionally useful for extracting the bending stiffness of tubes where one can measure stiffness at varying locations but cannot vary the effective thickness of the structure to verify the appropriate scaling for the bending rigidity. Examples might include not only macroscopic hollow cylinders as in the experiments shown here, but also biological micro-tubules \cite{DPSMS03}, carbon nano-tubes \cite{YKR00}, or rolled-up nanoparticle membranes \cite{WLMELJ15}, as long as these systems are sufficiently thin-walled and short. 


\begin{acknowledgments}
This work was supported by the Chicago MRSEC which is funded by NSF DMR-1420709. E.E. acknowledges support from the Simons foundation. Y.W. acknowledges support from NSF DMR-1508110.
\end{acknowledgments}

\bibliographystyle{unsrt}
\bibliography{scrols}
%

\end{document}